\documentclass[%
twocolumn,
 amsmath,amssymb,
 aps,
 showpacs
]{revtex4-1}

\usepackage{amsmath}
\usepackage{mathrsfs}
\usepackage{graphicx}% Include figure files
\usepackage{dcolumn}% Align table columns on decimal point
\usepackage{bm}% bold math
\usepackage{ulem}
\usepackage[%Uncomment any one of the following lines to test 
margin=.75in,
]{geometry}

\begin{document}
%\affiliation
\title{Effect of nuclear motion on high-order harmonic generation of H$_2^+$ in intense ultrashort laser pulses 
}

\author{Hamed Ahmadi$^{1}$}
\author{Ali Maghari$^{1}$}
\author{Hassan Sabzyan$^{2}$}
\author{Ali Reza Niknam$^{3}$}
\author{Mohsen Vafaee$^{4}$}
\email {m.vafaee@modares.ac.ir}

\affiliation{
$^{1}$Department of Physical Chemistry, School of Chemistry, College of Science, University of Tehran, Tehran, I. R. Iran
%\address{$^{3}$Department of Chemistry, University of Isfahan, Isfahan 81746-73441, I. R. Iran}
\\$^{2}$Department of Chemistry, University of Isfahan, Isfahan 81746-73441, I. R. Iran
\\$^{3}$Laser and Plasma Research Institute, Shahid Beheshti University, G.C., Tehran, I. R. Iran 
\\$^{4}$Department of Chemistry, Tarbiat Modares University, P. O. Box 14115-175, Tehran, I. R. Iran }

%\ead {m.vafaee@modares.ac.ir}
\begin{abstract}

High-order harmonic generation is investigated for H$_2^+$ and D$_2^+$ with and without Born-Oppenheimer approximation by numerical solution of full dimensional electronic time-dependent Schr\"{o}dinger equation under 4-cycle intense laser pulses of 800 nm wavelength and $I$=4, 5, 7, 10 $\times 10^{14}$ W$/$cm$^2$ intensities. For most harmonic orders, the intensity obtained for D$_2^+$ is higher than that for H$_2^+$, and the yield difference increases as the harmonic order increases. Only at some low harmonic orders, H$_2^+$ generates more intense harmonics compared to D$_2^+$. The results show that nuclear motion, ionization probability and system dimensionality must be simultaneously taken into account to properly explain the isotopic effects on high-order harmonic generation and to justify experimental observations.
\end{abstract}

%\vspace{2pc}
%\small{\textbf{Keywords:}  Time-dependent Schr\"{o}dinger equation, High-order harmonic generation, Beyond dipole approximation, Super-intense xuv ultrashort laser, Beyond Born-Oppenheimer approximation.}
%\submitto{\JPB}
%\twocolumn
%\title{}
\maketitle
%\twocolumn
\section{Introduction}
 High-order harmonic generation (HHG) is one of the phenomena observed in the interaction of intense laser pulses with atoms and molecules [1,2]. A three-step model for the description of the HHG mechanism has been proposed by Corkum [3] and extended by Lowenstein et al. [4]. In the first step of this mechanism, an electron wavepacket tunnels from an atom or a molecule into the continuum. In the second step, the electron moves away from the ion core, and after the field reverses, it is driven back to it. The third step arises when the electron recombines with its parent ion in which a high energetic photon is emitted. This model can be used in tunnelling regime which predicts maximum recollision energy of $3.17U_p$, where $U_p=I/{4\omega^2}$, is the pondermotive energy in which $I$ and $\omega$ are laser intensity and angular frequency, respectively. Based on the three-step model, for each harmonic order smaller than the cutoff harmonic order, we have two trajectories that contribute to the HHG with the same kinetic energy. These two trajectories return to their parent ion at different times. In one cycle of laser pulse, the electrons released over time interval $0.3 T_0<t<0.5 T_0$ ($T_0=2\pi/\omega_0$, with $\omega_0$ being laser frequency), return to the core during $0.5 T_0<t<0.95 T_0$. The trajectories travelled by these electrons are called short trajectories. While the electrons released over the $0.25 T_0<t<0.3 T_0$, return to the core during $0.95 T_0<t<1.25T_0$, and their path are called long trajectories because of longer round trip times than those of the short trajectories. The HHG is used to generate attosecond laser pulses [5,6] and to get structural information [7-11].

Here, we focus on the HHG reported for the H$_2^+$, H$_2$ and their corresponding isotopomers. The HHG process in molecules is more complex than that in atoms because of nuclear motion [12], two-center interference [13] and different orientations of molecule with respect to the laser field [13]. Effects of different initial vibrational states [14,15], initial nuclear velocities [16] and relation between electronic wavepacket expansion and internuclear distance on the HHG produced by H$_2^+$ [17] have been reported. In addition, extraction of nuclear dynamics from HHG spectra [18], effect of nuclear motion on the HHG efficiency by varying pump-probe time delay [19], on the broadening of cutoff regime  [20], on the length of generated attosecond laser pulses [21] and  on the generation of isolated attosencond laser pulses for H$_2^+$ have been investigated [22,23]. The HHG for H$_2$ and D$_2$ with higher yield for heavier isotopomer has been theoretically [12] and experimentally [8,24-25] reported. For most harmonic orders, experiments on H$_2$ and D$_2$ [8,24-25], and CH$_4$ and CD$_4$ [8] reveal the higher HHG yield for heavier isotopomers. While in theoretical works on H$_2^+$ and D$_2^+$, Feng \textit{et al}. [22] reported higher HHG yield in H$_2^+$ but Bandrauk \textit{et al}. [23] reported higher HHG yield for D$_2^+$.
In most theoretical works mentioned above, electron and nuclei are considered quantum mechanically in a one-dimensional (1D) model. The full-dimensional electron wavepacket expansion during laser interaction and its consequent HHG cannot be described properly by a one dimensional model [17]. 

In the present work, we are interested in the effect of the motion of nuclei on the HHG spectra by considering different isotopomers. As stated above, the theoretical results on the HHG yield on 1D H$_2^+$ and D$_2^+$ are different from experimental resulsts on H$_2$ and D$_2$. We want to address the discrepancy between these experimental and theoretical reports. The HHG is essentially a single-electron phenomenon and the HHG yield for H$_2^+$ and D$_2^+$ has not been experimentally reported because of difficulties in sample preparation. Therefore, in this work, besides comparing to theoretical works, we compare the HHG yield of different isotopomers of our single-electron systems with available experimental results obtained for two-electron systems. In this study, full-dimensional  electronic time-dependent Schr\"{o}dinger equation (TDSE) beyond Born-Oppenheimer approximation (NBO) is solved numerically for H$_2^+$, D$_2^+$ and X$_2^+$ (X is a virtual isotope of H being 10 times heavier). The full-dimensional electronic TDSE within Born-Oppenheimer approximation (BO) for H$_2^+$, which is indicated throughout the article by H$_2^+$(BO), is also solved to compare with NBO results. The equilibrium internuclear distance within BO is set to $R_e=1.96$ a.u. All calculations have been done with 4-cycle laser pulses of 800 nm wavelength with $I=$4, 5, 7 and 10 $\times 10^{14}$ W$/$cm$^2$ intensities. The Morlet-wavelet Fourier transform is used for time-frequency analysis of harmonics. We use atomic units throughout the article unless stated otherwise.

% $\vspace{-1cm}$
\section{Computational Methods}
The time-dependent Schr\"{o}dinger equation for H$_2^+$  (D$_2^+$) with electron cylindrical coordinate $(z,\rho)$ ͒ with respect to the molecular center of mass and internuclear distance $R$, for both $z$ and $R$ parallel to the laser polarization direction, can be read (after elimination of the center-of-mass motion) as [26-27]

\begin{eqnarray}\label{eq:1}
  i \frac{\partial \psi(z,\rho, R,t)}{\partial t}={\widehat H}(z,\rho, R,t)\psi(z,\rho, R,t).
\end{eqnarray}
In this equation, \^{H} is the total electronic and nuclear Hamiltonian which is given by
\begin{eqnarray}\label{eq:2}
 \widehat{H}(z,\rho, R,t)=&-\frac{2m_N+m_e}{4m_Nm_e}[\frac{\partial^2}{\partial \rho^2}+\frac{1}{\rho}\frac{\partial}{\partial \rho}+\frac{\partial^2}{\partial z^2}]
\nonumber \\
& -\frac{1}{m_N}\frac{\partial^2}{\partial R^2}+V_C(z,\rho, R,t),
\end{eqnarray}
with
\begin{eqnarray}\label{eq:3}
 \widehat{V}_C&(z,\rho, R,t)=-\frac{1}{\sqrt{(z+\frac{R}{2})^2+\rho^2}}-\frac{1}{\sqrt{(z-\frac{R}{2})^2+\rho^2}}
\nonumber \\
 &+\frac{1}{R}+(\frac{2m_N+2m_e}{2m_N+me})zE_0f(t)sin(\omega t).
\end{eqnarray}
In these equations, $E_0$ is the laser peak amplitude, $m_e$ and $m_N$ are the masses of electron and single nuclei, $\omega$ angular frequency  and \textit{f}(t) is the laser pulse envelope which is considered to have Gaussian form as
\begin{eqnarray}\label{eq:4}
  {\textit{f}}(t)=exp[-\frac{4 \ln(2)(t-2T_0)^2}{\tau^2}],
\end{eqnarray}
in which $\tau$ is a measure of the laser pulse duration (full width at half maximum (FWHM)) which is set to 3 femtosecond ($\sim$ 124 a.u.). The shape of the electric field of the laser pulse used in this work is shown in Fig. 1.

%===========================Figure===========================================
\begin{figure}[ht]
\begin{center}
\begin{tabular}{c}
\centering
\resizebox{70mm}{50mm}{\includegraphics{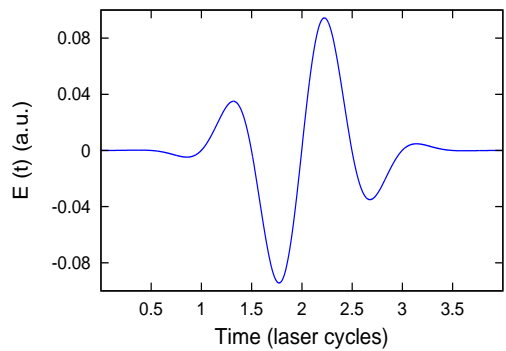}}
\end{tabular}
\caption{
\label{HHG} 
The shape of the laser electric field with a Gaussian envelope (Eq. (4)) with $\tau$=3 femtosecond ($\sim$ 124 a.u.) at 800 nm wavelength ($\omega=0.057$ a.u.) and $I$=4$\times 10^{14}$ W/cm$^2$ intensity.		}
\end{center}
\end{figure}
%====================================

The TDSE is solved using unitary split-operator methods [28-29] with 11-point finite difference scheme through a general nonlinear coordinate transformation for both electronic and nuclear coordinates which is described in more details in our previous works [30-32]. The grid points for z, $\rho$, and R coordinates are 300,
83, and 210, respectively. The finest grid size values in this
adaptive grid schemes are 0.13, 0.1, and 0.025, respectively for $z$, $\rho$, and $R$ coordinates. The grids extend up to $z_{max}
= 34$, $\rho_{max} = 25$, and $R_{max}= 16$. Only for $I$=1$\times 10^{15}$ W$/$cm$^2$ intensity, the $z$ grid points  is set to 500 and $z_{max}
= 98$. The size of the simulation boxes is considered large enough so that the loss of norm at the end of laser pulse does not exceed a few percent.
The HHG spectra are calculated as square of the windowed Fourier transform of dipole acceleration $a_z(t)$ in the electric field direction (z) as
\begin{eqnarray}\label{eq:5}
  S(\omega)= \vert \int_0^T a_z(t)\,H(t)\,exp[-i\omega t]\,dt\; \vert ^2,
\end{eqnarray}
where
\begin{eqnarray}\label{eq:6}
  H(t)= \frac{1}{2}[1-cos(2\pi \frac{t}{T})],
\end{eqnarray}
is the Hanning filter and $T$ is the total pulse duration which is set to 4 optical cycles (one optical cycle of 800 nm wavelength equals 2.6 fs). The time dependence of harmonics is obtained by Morlet-wavelet transform of dipole acceleration $a_z(t)$ via [33-34]
\begin{eqnarray}\label{eq:7}
  &w(\omega,t)= \sqrt{ \frac{\omega}{\pi^\frac{1}{2}\sigma}}\times
 \nonumber \\
 &\int_{-\infty}^{+\infty}a_z(t^\prime)exp[-i\omega (t^\prime-t)]exp[-\frac{\omega^2 (t^\prime-t)^2}{2\sigma^2}]dt^\prime.
\end{eqnarray}
We tried different $\sigma$ values and obtained the best time-frequency resolution by $\sigma=2\pi$ which is used in this work. Results of the calculations with and without Born-Oppenheimer (fixed-nuclei) approximation are denoted by BO and NBO, respectively.

\section{Results and Discussion}
The HHG spectra of H$_2^+$, D$_2^+$, X$_2^+$ and H$_2^+$(BO) obtained under 4-cycle 800 nm laser pulses of $I$=4, 5, 7, 10 $\times 10^{14}$  W$/$cm$^2$ intensities are shown in Fig. 2.
%===========================Figure===========================================
\begin{figure*}[ht]
%\begin{}
%\begin{tabular}{c}
\centering
%\captionsetup{justification=centering}
\resizebox{120mm}{110mm}{\includegraphics{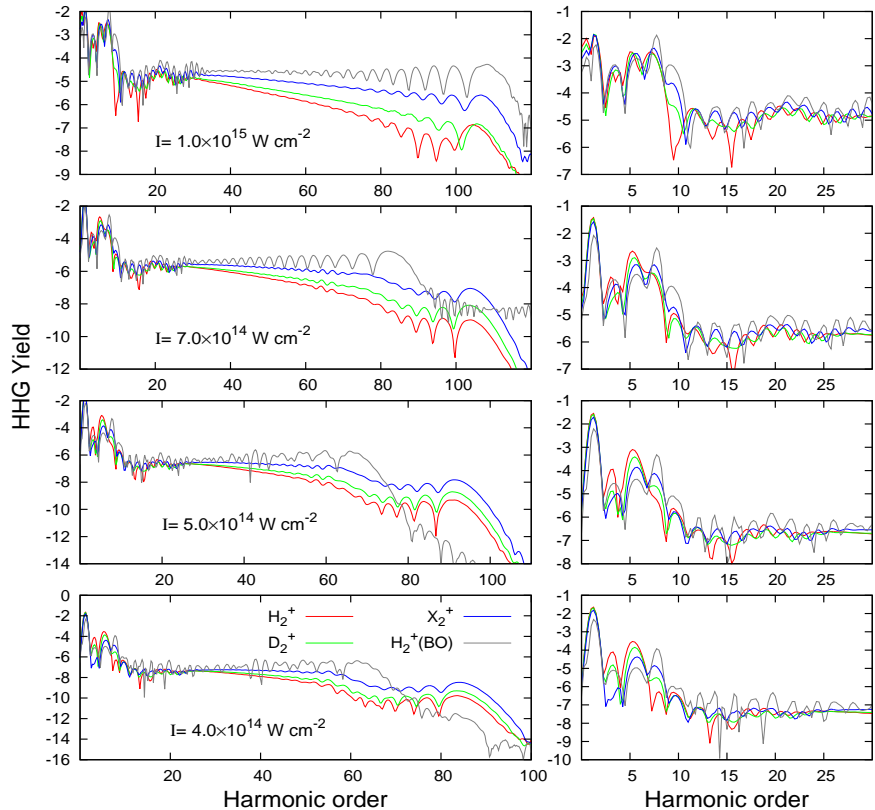}}
%\end{tabular}
\caption{
%\label{PIR}
High-order harmonic spectra produced by the NBO H$_2^+$, D$_2^+$, X$_2^+$  and BO H$_2^+$ (H$_2^+$(BO)) under 4-cycle laser pulses of 800 nm wavelength at $I=$4, 5, 7 and 10 $\times 10^{14}$ W$/$cm$^2$ intensities. For more clarity, the range of 0-30 harmonics of the spectra are magnified and plotted in the right side of each panel.
}
%\end{center}
\end{figure*}
%============================================
For some low harmonics, the HHG yield is higher for H$_2^+$ than the other two isotopomers and H$_2^+$(BO) (right panels of Fig. 2), but for harmonic orders greater than $\sim$ 25, the HHG yield is higher for H$_2^+$(BO), and for NBO cases, the difference between the HHG yield of lighter and heavier isotopomers increases as the harmonic order increases. Such behaviours have already been observed in recent experiments. The higher HHG yield for H$_2^+$ than D$_2^+$ for some low harmonic orders is observed in experiments on H$_2$ and D$_2$ at 1300 nm wavelength [25]. Higher HHG yield for D$_2^+$ than for H$_2^+$ at higher harmonic orders is also reported in experiments on H$_2$ and D$_2$ at 800 and 1300 nm wavelengths [8,24-25]. 

 As shown in Fig. 2, for the intensities used in our calculations, the cutoff occurs at harmonic orders smaller than that predicted by the three-step model [4]. The cutoff harmonic order $N_c$, according to the three-step model, for H$_2^+$, with ionization potential $I_p=1.1$ a.u., under 800 nm wavelength ($\omega_l=0.057$ a.u.) and different intensities are given in TABLE I.
 \begin{table}
 \caption{\label{Table}
The cutoff harmonic order $N_c$ according to the three-step model for H$_2^+$, with ionization potential $I_p=1.1$ a.u., and pondermotive energy $U_p$ under 800 nm wavelength ($\omega_l=0.057$ a.u.) and $I$=4, 5, 7, 10 $\times 10^{14}$ W$/$cm$^2$ intensities. The effective intensity (of the two central peaks, $I_e$) due to Gaussian envelope is also given for each laser intensity, $I$. The $N_c$ and $U_p$ are calculated for effective intensities.
} 
 \begin{tabular}{|c|c|c|}
\hline
 $I (I_e)$ W/cm$^2$& $U_p\, (a.u.)$ & $N_c=(3.17U_p+1.32I_p)/\omega_l$ \\ 
 \hline 
 4 (3.15)$\times 10^{14}$ & 0.69 & 64 \\ 
 \hline 
 5 (3.9)$\times 10^{14}$& 0.86 & 73 \\ 
 \hline 
 7 (5.5)$\times 10^{14}$& 1.21 & 93 \\ 
 \hline 
 10 (7.8)$\times 10^{14}$ & 1.71 & 121 \\ 
 \hline
 %\label{
%\label{PIR}
 \end{tabular} 
 \end{table}
  The plateau region is wider for heavier isotopomers and the NBO cutoff approaches the BO cutoff as the isotopomer becomes heavier.   
 For harmonic orders between 13-25 (Fig. 2), everywhere H$_2^+$ spectrum has valley (peak), the X$_2^+$ spectrum has peak (valley), and also as isotopomer becomes heavier, spectra modulation approaches to the corresponding BO spectra, as we see similar modulation between X$_2^+$ and H$_2^+$(BO). 
  
To justify the above observations, we study ionization probability, nuclear motion and compare these full-dimensional results with those reported for 1D models [22,23]. 
  First, the time-dependent contribution of ground state population in the evolving wavepacket for the four laser pulses with different intensities are calculated and demonstrated in Fig. 3. 
%===========================Figure===========================================
\begin{figure*}[ht]
\begin{center}
\begin{tabular}{c}
\centering
\resizebox{140mm}{90mm}{\includegraphics{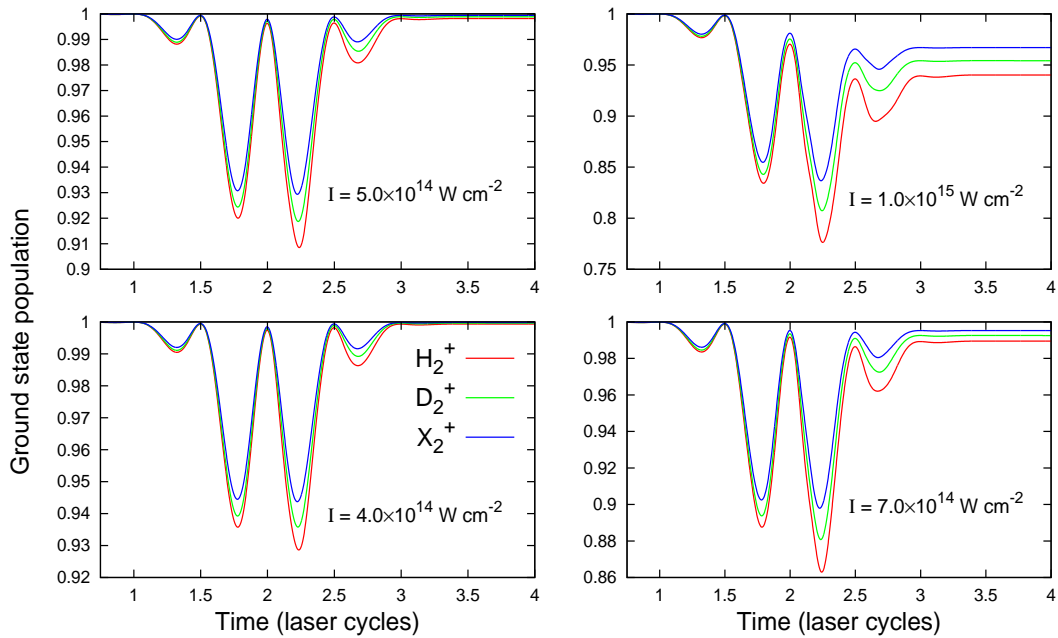}}
\end{tabular}
\caption{
\label{internuclear_distance}
The time-dependent contribution of the ground state to the evolving wavepacket of H$_2^+$, D$_2^+$ and X$_2^+$ exposed to 4-cycle laser pulses of 800 nm wavelength and $I$=4, 5, 7, 10 $\times 10^{14}$ W$/$cm$^2$ intensities.
		}
\end{center}
\end{figure*}
%==========================================
This figure shows that the ground state population is lower for H$_2^+$ than for D$_2^+$ and X$_2^+$. When ground state population is lower, higher ionization is expected. This higher ionization plays a positive role in the HHG enhancement for all harmonics because the more released electron results in more return of the released electron to the core giving rise to the HHG.  Therefore, the higher ionization can justify higher HHG yield observed for H$_2^+$ than those observed for other isotopomers at low harmonic orders. 

To demonstrate better the effects of nuclear motion on the HHG spectra, the Morlet-wavelet Fourier transform of the HHG spectra of Fig. 2 for BO and NBO cases, and the time-dependent average internuclear distance for NBO cases are derived and depicted in Figs. 4 and 5, respectively. 
%===========================Figure===========================================
\begin{figure*}[ht]
\begin{center}
\begin{tabular}{c}
\centering
\resizebox{140mm}{110mm}{\includegraphics{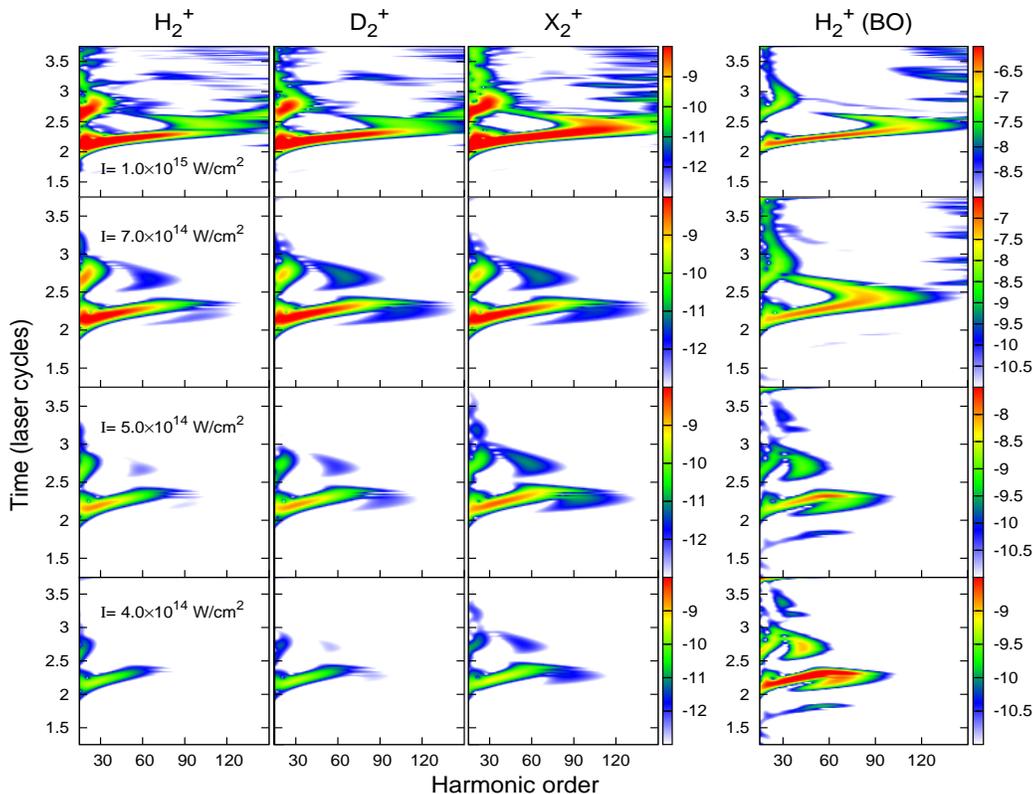}}
\end{tabular}
\caption{
\label{norm}
The Morlet-wavelet time profiles of dipole acceleration for NBO H$_2^+$, D$_2^+$ and X$_2^+$ and BO H$_2^+$ (H$_2^+$(BO)) under 4-cycle laser pulses of 800 nm wavelength and $I$=4, 5, 7, 10 $\times 10^{14}$ W$/$cm$^2$ intensities. The HHG intensities are depicted in color scales on the right side of panels.
		}
\end{center}
\end{figure*}
%=================================
Each row of Fig. 4 is related to a specific (labelled) intensity. For all intensities in Fig. 4, the three peaks are observed around 1.75, 2.25 and 2.75 optical cycles (o.c.) which, based on the three-step model, correspond to electron trajectories released at 1.325-1.5, 1.775-2 and 2.225-2.5 o.c., respectively. The very weak peak around 1.75 o.c., which is only present for $I$=4 and 5 $\times 10^{14}$ W$/$cm$^2$ intensities with BO, is related to the short trajectories born during 1.325-1.5 o.c. which is suppressed when NBO is considered as a result of nuclear wavefunction spreading that is justified in Ref. [23]. For the weak peak around 2.75 o.c., we see almost both short and long trajectories which are limited to low harmonic orders not of importance in this work. The strong peak around 2.25 o.c. is the most important peak which is extended to high harmonic orders and we concentrate on it. Note that for harmonic orders greater than $\sim$ 30, this peak is only responsible for differences seen and stated for Fig. 2. The birth and return times of this peak are between 1.775-2.675 o.c. in which the electric field has two strong minimum (at 1.775 o.c.) and maximum (at 2.225 o.c.). The electric field strength should be high enough to ionize and accelerate the electron, and to return the released electrons to the core with higher energy. It can be said that for most panels in Fig. 4, the short-trajectory branch  dominantly contribute to the peak around 2.25 o.c., and the weak long-trajectory branch is seen only for H$_2^+$(BO) at $I$=7 $\times 10^{14}$ W$/$cm$^2$ intensity.
  
As shown in Fig. 5, the internuclear distance is increased more for the lighter isotopomer H$_2^+$ than D$_2^+$ and X$_2^+$. This increase is higher for higher laser intensities. The sharp increase in the internuclear distance of lighter isotopomer can affect the HHG spectra. We can see the effect of nuclear motion in two cases: harmonic orders 13-25 and above 25. For 13-25 harmonic orders, opposite modulation of X$_2^+$ and H$_2^+$ HHG spectra and similar modulation of X$_2^+$ and H$_2^+$(BO) HHG spectra (Fig. 2), suggest that nuclear motion is responsible for these observations. Note that electronic structures of different isotopomers used in this work are similar and therefore the difference between their HHG spectra can be attributed to their nuclear motion only. It should, however, be mentioned that ionization rate of these isotopomers are different and we showed that H$_2^+$ has higher ionization than other (Fig. 3). Now, we consider nuclear motion for harmonic orders greater than 25. As stated before, there is only one peak around 2.25 o.c which is mainly related to the short trajectories that contribute to the HHG for high harmonic orders. As it can be deduced from Fig. 4 by looking at short-trajectory peaks, higher harmonic orders are produced at longer times (the time between the birth and recollision of the ionized electron increases as harmonic order increases) which is vice versa for long trajectories [33]. The HHG attenuation of H$_2^+$ compared to those of heavier isotopomers and H$_2^+$(BO) can be related to more increase in the internuclear distance for lighter isotopomer (Fig. 5). The difference between the HHG yield of H$_2^+$ and those of other isotopomers is increased with the increase of the harmonic order because higher harmonic orders are produced at longer times when there is a larger internuclear distance for lighter isotopomer, giving rise thus to attenuation of the HHG.

For all intensities in Fig. 4, the HHG produced in the long trajectory is very weak compared to that of the short trajectory. For low intensities, $I$=4 and 5 $\times 10^{14}$ W$/$cm$^2$, long trajectory is completely suppressed for both BO and NBO cases. As shown in Fig 4, the long-trajectory attenuation is more obvious for NBO than for BO, and for lightest isotopomer is more distinct because of faster and larger nuclear motion. The long-trajectory suppression in NBO has already been observed for the 1D electronic NBO calculations on H$_2^+$ [23]. Nevertheless, the 1D BO results for which short and long trajectories have similar HHG strengths [23], are not in agreement with the long-trajectory suppression observed with full-dimensional electronic BO calculations carried out in this work. This difference is because of the drift induced in the laser propagation direction, which can not happen in 1D models, and thus reduces the recollision of electrons in different trajectories into their parent ions.

%===========================Figure===========================================
\begin{figure*}[ht]
\begin{center}
\begin{tabular}{l}
\resizebox{140mm}{90mm}{\includegraphics{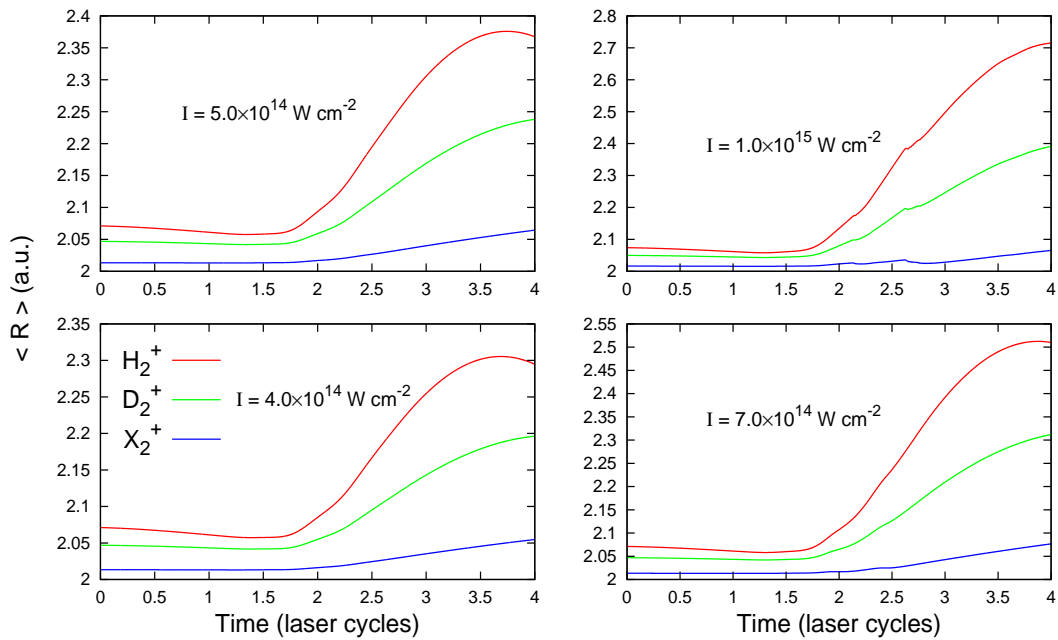}}
\end{tabular}
\caption{
\label{<z>}
The same as FIG. 3, but for time-dependent internuclear distance.		
		}
\end{center}
\end{figure*}
%===================================

 As stated before, for both NBO and BO cases, the cutoff position is lower than that obtained by the three-step model and 1D NBO calculations [22]. Two reasons can be considered here. First, taking into account electron in full dimension causes electron with large return time to lose its chance to recollide with its parent ion (the electron drifts in the propagation direction). Second, nuclear motion also leads to the suppression of the harmonics near the cutoff. Since the short trajectories have higher contributions to the HHG spectra, especially near the cutoff, and for these trajectories, higher harmonics are produced at longer times, and at longer times, the nuclei are more displaced away from their equilibrium positions, the role of nuclear motion in attenuating the HHG spectra is increased with increasing harmonic order. This can also be deduced when the NBO cutoff approaches the BO cutoff as the isotopomer becomes heavier (Fig. 2).        

For most harmonics (except for the first few low harmonic orders), the HHG is more intense for heavier isotopomers, which is in agreement with experimental reports [8,24-25] and is in contrast to the results predicted by the 1D electronic NBO calculation, that is the lighter isotopomer produces higher HHG yield over the whole spectrum [22]. Feng \textit{et al}. explained their results based on the ionization probabilities which is valid only for 1D NBO calculations [22]. Based on the full-dimensional electronic NBO calculations carried out in this work, it can be stated that more ionization occurs for lighter isotopomers which is responsible for higher HHG yield at low harmonic orders. In 1D electronic model, the recollision is overestimated compared to that in the real full-dimensional calculations. The recollision-recombination occurrence is suppressed for full-dimensional electronic NBO cases because of drift induced in the laser propagation direction and the increase of internuclear distance.

\section{Conclusion}
 We solved numerically time-dependent Schr\"{o}dinger equation for H$_2^+$ and D$_2^+$ with and without Born-Oppenheimer approximation to investigate the effect of the nuclear motion via analysis of the high-order harmonic generation to address the discrepancy between the HHG yield obtained for H$_2^+$ and D$_2^+$ isotopomers by one-dimensional non-Born-Oppenheimer calculations [22,23] and those experimentally observed on H$_2$ and D$_2$ [8,24-25]. While, our results show that when nuclear motion is taken into account, higher HHG yield is obtained for heavier isotopomer which is compatible with experimental reports [8,24-25]. The 1D electronic NBO calculations overestimate the recollision-recombination effect relative to full-dimensional electronic NBO calculations, especially at longer return times of the released electron corresponding to increased internuclear distance, which is detrimental to the recollision-recombination phenomenon, and thus attenuates the HHG production. 
%=====================================Refrences=======
\section{References}
\bibliography{p7}% Produces the bibliography via BibTeX.

\begin{thebibliography}{100}
\bibitem{1} 
T. Brabec and F. Krausz, Rev. Mod. Phys. \textbf{72}, 545 
(2000). 

%
\bibitem{2}
C. Winterfeldt, C. Spielmann, and G. Gerber, Rev. Mod. Phys. 
\textbf{80}, 117 (2008). 
 
\bibitem{3}
 P. B. Corkum, Phys. Rev. Lett. \textbf{71}, 1994 (1993).

\bibitem{4}
M. Lewenstein, P. Balcou, M. Y. Ivanov, A. L$^{^,}$Huillier and P. A. Corkum, Phys. Rev. A \textbf{49}, 2117 (1994).

\bibitem{5}
E. Goulielmakis, V. S. Yakovlev, A. L. Cavalieri, M. Uiberacker,
V. Pervak, A. Apolonski, R. Kienberger, U. Kleineberg, and
F. Krausz, Science \textbf{317}, 769 (2007).

\bibitem{6}
F. Krausz and M. Y. Ivanov, Rev. Mod. Phys. \textbf{81}, 163 (2009).


\bibitem{7}
J. Itatani, J. Levesque, D. Zeidler, H. Niikura, H. P\'{e}pin, J. C.
Kieffer, P. B. Corkum, and D. M. Villeneuve, Nature (London)
\textbf{432}, 867 (2004).

\bibitem{8}
S. Baker, J. S. Robinson, C. A. Haworth, H. Teng, R. A.
Smith, C. C. Chiril$\breve{a}$, M. Lein, J. W. G. Tisch, and J. P.
Marangos, Science \textbf{312}, 424 (2006).

\bibitem{9}
S. Haessler, J. Caillat, W. Boutu, C. Giovanetti-Teixeira,
T. Ruchon, T. Auguste, Z. Diveki, P. Breger, A. Maquet,
B. Carr\'{e}, R. Ta\"{i}eb, and P. Sali\`{e}res, Nat. Phys. \textbf{6}, 200 (2010).

\bibitem{10}
C. Vozzi, M. Negro, F. Calegari, G. Sansone, M. Nisoli, S. D.
Silvestri, and S. Stagira, Nat. Phys. \textbf{7}, 822 (2011). 

\bibitem{11}
 W. Li, X. Zhou, R. Lock, S. Patchkovskii, A. Stolow, H. C.
Kapteyn, and M. M. Murnane, Science \textbf{322}, 1207 (2008). 

\bibitem{12}
M. Lein, Phys. Rev. Lett. \textbf{94}, 053004 (2005).

\bibitem{13}
M. Lein, N. Hay, R. Velotta, J. P. Marangos, and P. L.
Knight, Phys. Rev. Lett. \textbf{88}, 183903 (2002).


\bibitem{14}
  Y. H. Guo, H. X. He, J. Y. Liu and G. Z. He, J. Mol. Struct. \textbf{947}, 119 (2010).

\bibitem{15}
 J. Zhao and Z. Zhao, Phys. Rev. A \textbf{78}, 053414 (2008).

\bibitem{16}
 N. T. Nguyen, V. H. Hoang and V. H. Le, Phys. Rev. A \textbf{88}, 023824 (2013).

\bibitem{17}
A. D. Bandrauk and H. Yu, J. Phys. B: At. Mol. Opt. Phys. \textbf{31}, 4243 (1998).

\bibitem{18}
 R. Daniele, G. Castiglia, P. Corso, E. Fiordilino, F. Morales and G. Orlando, J. Mod. Opt. \textbf{56}, 751 (2009).

\bibitem{19}
T. Bredtmann, S. Chelkowski and A. D. Bandrauk, J. Phys. Chem. A \textbf{116}, 11398 (2012).

\bibitem{20}
W. Qu, Z. Chen, Z. Xu and C. H. Keitel, Phys. Rev. A \textbf{65}, 013402 (2001).

\bibitem{21}
A. D. Bandrauk, S. Chelkowski and H. Lu, J. Phys. B: At. Mol. Opt. Phys. \textbf{42}, 075602 (2009).

\bibitem{22}
L. Feng and T. Chu, J. Chem. Phys. \textbf{136}, 054102 (2012).


\bibitem{23}
X. L. Ge, T. Wang, J. Guo, and X. S. Liu Phys. Rev. A \textbf{89}, 023424 (2014).

\bibitem{24}
S. Baker, J. S. Robinson, M. Lein, C. C. Chiril$\breve{a}$, R. Torres, H. C.
Bandulet, D. Comtois, J. C. Kieffer, D. M. Villeneuve, J. W. G.
Tisch, and J. P. Marangos, Phys. Rev. Lett. \textbf{101}, 053901 (2008).

\bibitem{25}
H. Mizutani, S. Minemoto, Y. Oguchi and
H. Sakai, J. Phys. B: At. Mol. Opt. Phys. \textbf{44},  081002 (2011).



\bibitem{26}
J. R. Hiskes, Phys. Rev. \textbf{122}, 1207 (1961). 


\bibitem{27}
M. Vafaee, Phys. Rev. A \textbf{78}, 023410 (2008).

\bibitem{28}
A. D. Bandrauk and H. Shen, J. Chem. Phys. \textbf{99}, 1185 (͑1993)͒.

\bibitem{29}
 M. D. Feit, J. A. Fleck, Jr. , and A. Steiger, J. Comput. Phys. \textbf{47}, 412 (1982).
 
\bibitem{30}
M. Vafaee and H. Sabzyan, J. Phys. B \textbf{37}, 4143 (2004).

\bibitem{31}
M. Vafaee, H. Sabzyan, Z. Vafaee, and A. Katanforoush,
e-print arXiv:physics/0509072.


\bibitem{32}
M. Vafaee, H. Sabzyan, Z. Vafaee, and A. Katanforoush,
Phys. Rev. A \textbf{74}, 043416 (2006).

\bibitem{33}
C. Chandre, S. Wiggins and T. Uzer, Phys. D \textbf{181}, 171 (2003).

\bibitem{34}
A. D. Bandrauk, S. Chelkowski and H. Lu, Chem. Phys. \textbf{414}, 73 (2013).

\end{thebibliography}

\clearpage
\end{document}